\documentclass[twocolumn,aps,amsmath,superscriptaddress,prl]{revtex4}
\usepackage{graphicx,dcolumn,bm,color}
\usepackage{dsfont}
\usepackage{amsfonts}
\usepackage{amssymb}
\usepackage{dcolumn}
\usepackage{bm}

\usepackage{amsmath} 
\usepackage[latin1]{inputenc}
\DeclareMathAlphabet{\mathpzc}{OT1}{pzc}{m}{it}

\begin{document}

\title{Chiral molecular films as electron polarizers and polarization modulators}

\author{Ernesto Medina}
\affiliation{Laboratorio de F\'isica Estad\'istica de Sistemas
Desordenados, Centro de F\'isica, IVIC, Apartado 21827, Caracas 1020A, Venezuela.}

\author{Floralba L\'opez}
\affiliation{Quimicof\'isica de Fluidos y Fen\'omenos Interfaciales (QUIFFIS), Departamento de Qu\'imica,
Universidad de los Andes, M\'erida 5101, Venezuela.}

\author{Mark A. Ratner}
\affiliation{Department of Chemistry, Northwestern University, 2148 Sheridan Rd. Evanston, IL 60208, USA}

\author{Vladimiro Mujica}
\affiliation{Department of Chemistry, Northwestern University, 2148 Sheridan Rd. Evanston, IL 60208, USA}
\affiliation{Department of Chemistry and Biochemistry, Arizona State University, Tempe, AZ 85287 USA.}
\affiliation{Center for Nanoscale Materials, Argonne National Laboratory, Argonne IL 60439, U.S.A.}

\date{\today}

\begin{abstract}
Recent experiments on electron scattering through molecular films have shown that
chiral molecules can be efficient sources of polarized electrons even in the absence
of heavy nuclei as source of a strong spin-orbit interaction. We show that self-assembled monolayers (SAMs)
of chiral molecules are strong electron polarizers due to the high density effect of the monolayers and
explicitly compute the scattering amplitude off a helical molecular model of carbon atoms. Longitudinal 
polarization is shown to be the signature of chiral scattering. For elastic scattering, we find that at 
least double scattering events must take place for longitudinal polarization to arise. We predict energy 
windows for strong polarization, determined by the energy dependences of spin-orbit strength and multiple scattering
probability. An incoherent mechanism for polarization amplification is proposed, that increases the
polarization linearly with the number of helix turns, consistent with recent experiments on DNA SAMs.

\end{abstract}
\maketitle

The surprising experimental finding that self assembled monolayers (SAMs) of chiral organic molecules on a gold substrate can act as highly efficient spin filters or polarizers, with a polarization factor larger than 60\%\cite{NaamanDNA}, has prompted us to examine the underlying physical mechanism responsible for this behavior. In the absence of external fields, spin-dependent effects in electronic transport across an interface between two materials are essentially determined by electron-electron correlation and spin orbit (SO) interaction\cite{DasSarma}.  In the experiments we are referring to, photoelectrons emitted from the gold surface are transmitted through the SAM at energies larger than the interfacial tunneling barriers\cite{Naaman1999,NaamanDNA}. Under these conditions electron transmission is mostly a one-particle process and therefore the explanation for the spin polarization effects must physically depend on either the presence of strong internal magnetic fields or the spin-orbit interaction.  We argue in this letter that the most striking experimental observations: the role of chirality and the prevalence of longitudinal polarization, can be understood from a basic principles model using a one particle hamiltonian that includes the bare spin-orbit interaction. 

Spin-dependent electron scattering has been observed in gas phase experiments\cite{Kessler96} with chiral molecules as targets, and the observed asymmetry factor for spin polarization can be well understood as due to the result of a substantial spin-orbit interaction caused by a heavy atom present in the molecule. A similar explanation can be attempted for the experiments with SAMs\cite{Yeganeh}, but the objection has been raised that for the low atomic number atoms, C, H, N and O present in the monolayers, the magnitude of the SO interaction is too small to explain the observed polarization\cite{XieNanoLett}. In a very recent experiment reported by G\"ohler et al\cite{NaamanDNA}, a direct measurement of spin polarization with a strong longitudinal polarization was obtained for an unpolarized emerging electron beam. This experiment establishes in a definitive way that the SAM, composed of light atoms only, can polarize electrons regardless of the initial origin and magnitude of the electron polarization.
\begin{figure}
\includegraphics[width=7.5cm]{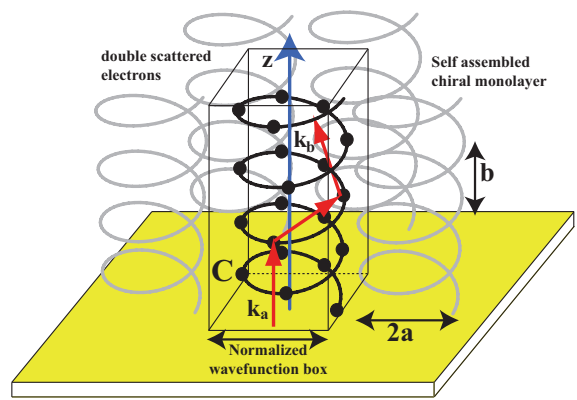}
\caption{Scattering model of a chiral assembly of molecules chemisorbed on a gold surface. Chiral molecules are assumed helices of radius $a$ and pitch $b$ built from carbon atoms, whose electronic state is fully described in the model. The electrons are incident in the z direction and are assumed to be scattered internally within the layer, and then received at a Mott detector. Electron wavefunctions from surface are normalized within a rectangular box of volume $v$.}
\label{fig1}
\end{figure}
We use a model for elastic spin-dependent multiple scattering through chiral molecules to explain these experimental findings. Three key ingredients emerge: (i) Although transverse polarization can arise under inelastic events from point targets, the appearance of longitudinal polarization requires the existence of chirality. (ii) Strongly enhanced spin orbit interaction is due to the fact that the density of molecules in a SAM is comparable to that of a solid structure, thereby increasing the electron wavefunction overlap with deeply penetrating orbitals as compared to the gas phase\cite{Kessler96}. (iii) Multiple scattering is needed to produce longitudinal polarization, so that there is an energy range where electron polarization is most effective in the few electron volts range, as seen in the experiments. We claim that sources of spin-orbit coupling analysed\cite{Gutierrez} such as electric polarization fields, even though as large as $10^6-10^7$V/cm\cite{Cahen}), cannot account for the spin-splitting energy necessary for strong spin polarization. Such of SO also do not carry the field configuration to transfer the appropriate longitudinal angular momentum to the transmitted electrons. It is the wavefunction penetration to the nucleus in atomic orbitals that will yield electric fields of $10^{10}$V/cm at distances of a Bohr orbit within the atom. This is the source of the peculiar strength of SO in non-centrosymmetric semiconductors\cite{Winkler}. 

The detailed calculation of the scattering cross section and polarization of even a small molecule like ${\rm H}_2{\rm O}$ \cite{GreerThompson} can be very demanding, taking into account static, exchange and spin-orbit interactions in detail. In order to capture the coherence effects by connecting them to molecular parameters, it is necessary to have analytical solutions to simpler models. In this paper we solve explicitly for the scattering of electrons from a single oriented helical molecule\cite{XieNanoLett}, in a SAM as depicted in Fig.\ref{fig1}, in the Born double scattering approximation. We consider the elastic scattering of spin-quantized electrons from a chiral potential with explicit consideration for the SO interaction and scalar potential effects from the particular atomic centers that conform the molecular structure. We solve for the transition amplitude in the scattering formulation where electrons are incident with relative momentum $\bm k$, and are subject to the potential $V({\bf r},\bm\sigma)=V({\bf r}){\mathds 1} +\alpha\hat{\bm\sigma}. \left[\nabla V({\bf r})\wedge{\bf p}\right]$, at position ${\bf r}$, including both scalar and spin dependent contributions with $\alpha=-\frac{\hbar}{(2m_e c)^2}$ the bare Pauli SO prefactor and $V({\bm r})$ the atomic Coulomb potential. 

In order to compute the scattering amplitudes, we need to expand the $\bf t$ matrix in powers of the potentials involved. In the double scattering approximation, and assuming non-overlapping multicentral potentials\cite{MottMassey}, we have that for a multiatomic molecule
assumed to be composed of carbon atoms $C^{(1)}$ through $C^{(6)}$
\begin{eqnarray}
{\hat {\bf f}}({\bf k}_a,{\bf k}_b)&=&-\left (\frac{2 \pi m}{h^2}\right ) \Big (\sum_l^{\rm helix} \langle k_b| t_{C_{(l)}} | k_a\rangle\label{MultiCentral}\\
&+&\sum_{l\ne m}^{\rm helix} \langle k_b| t_{C^{(l)}}G_0 t_{C^{(m)}} | k_a\rangle+......\Big ),\notag
\end{eqnarray}
which includes both single and double scattering contributions. Here, ${\bf k}_a,~{\bf k}_b$ are the scattering incoming/outgoing wavevectors. The subindices of the single scattering $t$ matrix correspond to the particular atom involved $C^{(i)}$ and $G_0=\exp(ik|{\bf r}_A-{\bf r}_B|)/4\pi|{\bf r}_A-{\bf r}_B|$ for atomic centers $A$ and $B$. The single scattering $t_A$ matrix elements are given by
$\langle {\bf k}_o| t_A|{\bf k}_i\rangle =-\frac{h^2}{2 \pi m}\exp(i({\bf k}_i-{\bf k}_o)\cdot{\bf r}_A)\hat{\bf f}_A({\bf k}_i,{\bf k}_o)$
with the single scattering amplitude
\begin{eqnarray}
\hat{\bf f}_A({\bf k}_i,{\bf k}_o)&=&-\frac{\mu}{2\pi \hbar^2}\int \frac{\text{e}^{-i{\bf k}_o .{\bf r}}}{\sqrt{v}} V({\bf r},\hat{\sigma})\frac{\text{e}^{i{\bf k}_i . {\bf r}}}{\sqrt{v}}  d^3{\bf r} \notag \\
       &=&-\Gamma({\bm k}_i,{\bm k}_o)
       \left[ \mathds{1}+i\alpha \hbar\hat{\bm\sigma}. \left({\bm k}_o\wedge {\bm k}_i\right )\right ],
       \label{eqnFms}
\end{eqnarray}
where ${\bf k}_o,{\bf k}_i$ stand for collision input and output wavectors between any pair of atoms. The potentials have been translated to the origin, as their phase factors have been extracted from the integral. The denominator $v$ represent a volume per molecule where the incident electron wavefunction is normalized, and $\Gamma({\bm k}_i,{\bm k}_o)$ is the expectation value of the atomic potential in the plane-wave basis.  

Note a strong result from the above expression; no longitudinal polarization can result from single scattering events no matter the symmetry of the potential since only transverse components can arise (Pauli matrix components in the scattering plane). Nevertheless, double scattering events that result from products of single scattering amplitudes will contain all polarizations, as a result of the algebra of the Pauli matrices, physically controlling the transfer of orbital to spin angular momentum. Thus, one needs to go beyond single scattering to gain longitudinal polarization. Higher order scattering events, as we will see, have energy requirements that should be reflected in polarization experiments and will serve to prove their role. We also argue that higher order scattering is the only way to get chirality imprinted on the longitudinal spin signature. In view of the previous points, we consider the molecular potentials as a superposition of their atomic nuclei, disregarding bonding effects\cite{DillDehmer}. We employ the wavefunctions of Clementi and Roetti\cite{Clementi}, derived from a Hartree-Fock (HF) model, as convenient expressions to simplify explicit calculations.

The situation of highly packed SAMs can be compared to that of semiconductors, in the sense that the major part of the SO interaction strength comes from probing electric fields close to the atomic cores. The SO retains the strength associated with individual atoms in the solid\cite{Winkler}, the spin splitting energy being very similar (0.006 eV for carbon\cite{phillips}).  We can correctly estimate the SO interaction, {\it adjusting no parameters} i.e. from the microscopic Hamiltonian, by calculating the single particle amplitude as follows: We normalize the electron incident plane wave in a box whose volume is equivalent to the volume per molecule in the SAM. We then expand the plane waves in the local atomic wavefunction basis; $\Gamma({\bm k}_i,{\bm k}_o)=\sum_{nl,n'l'}\langle{\bm k}_o |\varphi_{nl}\rangle \langle\varphi_{nl}|V(r)|\varphi_{n'l'}\rangle\langle\varphi_{n'l'}|{\bm k}_i\rangle$, with $|\varphi_{nl}\rangle$ the HF wavefunctions and $|\bm k_{i,o}\rangle$  the input-output normalized plane wave vectors, and $V(r)$ the bare Coulomb potential. This estimate has to be done for each incoming/outgoing wavevector combination, to properly account for momentum transfer and adding over
the available orbitals of the target atoms which for carbon include $2p$ and higher levels with $l\ne0$ that can contribute to the SO scattering strength.  This procedure is a very simple and robust approach that accounts for the correct spin splitting and the density effect of the scattering targets, and for the order of magnitude differences between gas phase polarization\cite{Kessler96} and that resulting from electron scattering in the SAMs.

It is useful to rewrite the full scattering amplitude (one and two body scattering Eq. \ref{MultiCentral}) in the form
\begin{eqnarray}
\hat{\bf f}({\bf k}_a,{\bf k}_b)&=& -\frac{\mu}{2\pi \hbar^2}\big [h_0({\bf k}_a,{\bf k}_b){\mathds 1}+ h_1({\bf k}_a,{\bf k}_b)\sigma_x\notag \\ 
&+& h_2({\bf k}_a,{\bf k}_b) \sigma_y+h_3({\bf k}_a,{\bf k}_b)\sigma_z\big ],\notag\\
\label{finSigmas}
\end{eqnarray}
where the functions $h_i({\bf k}_a,{\bf k}_b)$ are to be determined. The scattered polarization component $P_l$, when the incident state is a completely unpolarized state $\rho=(1/2)\mathds{1}$, can be computed from the output density matrix ${\bm \rho'}={\bf f}{\bm \rho}{\bf f}^{\dag}$ through the relation $\text{Tr}({\bm \sigma} {\bm \rho'})/\text{Tr}{\bm \rho'}$. In terms of the expansion in Eq.\ref{finSigmas}, the polarization can then be written as
\begin{equation}
P_l = \frac{2(\Re(h_l h^{\dagger}_0)+\Re(i h_m h^{\dagger}_n))}{|h_0|^2+|h_1|^2+|h_2|^2+|h_3|^2}, \\
\label{polarization}
\end{equation}
where the indices $(l,m,n)$ are taken in cyclic order and $\Re$ denotes the real part of the argument. 
We immediately note a central feature of polarization effects in electron scattering: It is an interference effect between the spin-orbit interaction $h_i$ with $i\ne 0$ and the purely potential term $h_0$. One needs the presence of both to guarantee polarization. 
\begin{figure}
\includegraphics[width=8cm]{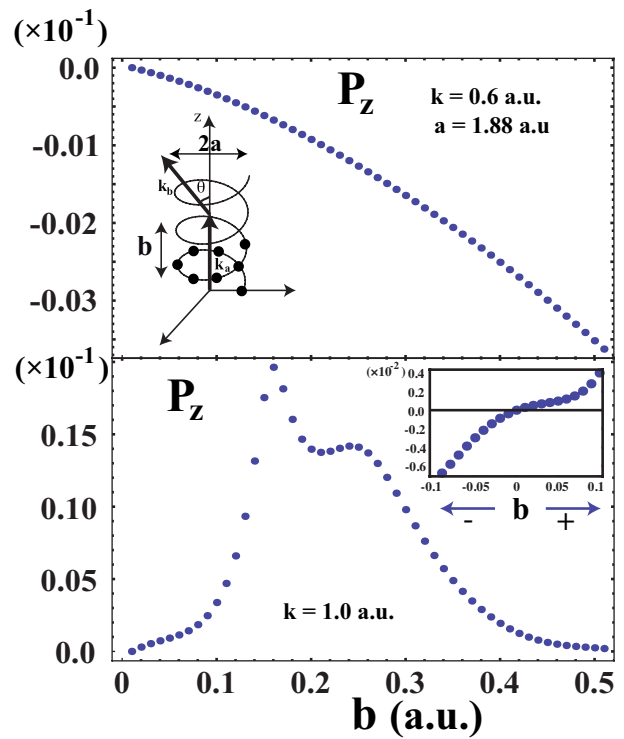}
\caption{Total longitudinal polarization per molecule, averaged over outgoing scattering angles, versus helix pitch $b$ as defined in the inset, for representative values of $k$. No longitudinal polarization is obtained for $b=0$ (chirality zero) for any value of the wavevector. The polarization is a non-monotonic function of $b$ beyond a certain pitch. The inset  (bottom panel) shows the change of the sign of the longitudinal polarization when chirality changes sign.}
\label{fig3}
\end{figure}
\begin{figure}
\includegraphics[width=7.5cm]{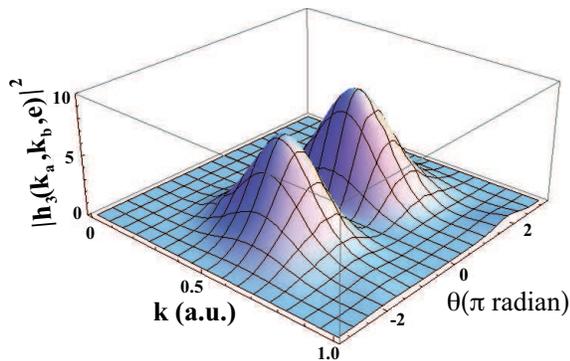}
\caption{The unnormalized probability amplitude for longitudinal polarization as a function of the incident wavevector and angular dependence. The figure shows the window of wavectors (in figure between k=0.3 and 1 a.u.) where longitudinal polarization is observed.}
\label{fig5}
\end{figure}

In Fig.\ref{fig3} we depict the electron polarization as a function of the pitch $b$ of the discrete helix built from six carbon atoms, that comprise a single turn of the structure. The range of $k$ values considered is between $0$ and $1$ a.u. (energies between 0 and 10 eV), in the range that has been explored in polarization experiments\cite{NaamanDNA}. For all values of the incident $k$, the longitudinal polarization vanishes as chirality, as measure by the pitch $b$, goes to zero. The relation between longitudinal polarization and pitch constitutes a measure of chirality itself, although the measure is obviously non-monotone, since there are important interference effects at play. The longitudinal polarization, can be of any sign depending on the range of wavevectors considered\cite{Kessler96} and the magnitude of the pitch of the scattering target. Note that one can achieve polarizations of 1\% for $k=1$ a.u. which is three orders of magnitude larger than gas phase polarizations reported with heavy atoms\cite{Kessler96}, and compatible with a comparable molecule in a SAM, such as Lysine\cite{Naaman1999}. A molecule such as DNA contains a much larger number of atoms per turn and a double helix.  It is beyond the scope of our approach to include such details in our model. 

To assess the energy dependence of the polarization, the best measure is the scattering probability in terms of the components defined by Eq.\ref{finSigmas}, since polarizations and asymmetries always yield a relative quantity to the total scattered intensity. In Fig.\ref{fig5} we see $|h_3|^2$ i.e. the amplitude of the longitudinal polarization, as a function of the incident wavevector $k$ and scattering angle $\theta$.  One can readily see that the relevant energy regime for chirality effects, in the double scattering approximation, are in the units of eV range (in a.u. in the figure). Since longitudinal polarization requires at least double scattering, the electrons must be deflected at large angles for intermediate dispersion events. Nevertheless, large deflections can only occur with the denominator handicap $\propto |{\bm k}_o-{\bm k}_i|^3$ the momentum transfer. If $k$ is increased by a factor of two, the spin-orbit interaction effects can be reduced by an order of magnitude, concomitantly reducing longitudinal polarization. There is a corresponding effect at lower $k$ where the spin-orbit interaction decreases linearly. So there will be a peak range in energies where longitudinal polarization is observed.

Up to this point we have only computed the polarizing power of a single turn of the helix under double scattering conditions. Successive turns of the helix or additional layers of chiral molecules can be treated incoherently by assuming layers of one helix turn thick. Since all scattering occurs at room temperature in the experiments, one does not expect long coherence lengths beyond a few nanometers. The scattering is then presumed to proceed by successive incoherent double scatterings through different turns of the helix. This can involve either the same helix or the ones within the lateral distance defined by the transverse component of the wavevector of the electrons emitted from the surface. 

\begin{figure}
\includegraphics[width=8.0cm]{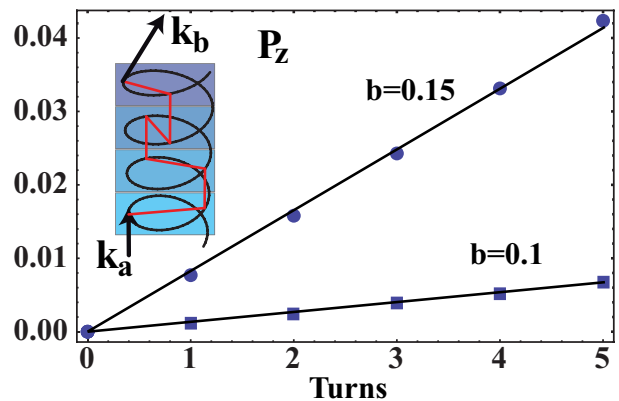}
\caption{The longitudinal polarization as a function of the number of turns in the helix for $b<a=1.88$ a.u. and $k=1~$a.u. The polarization grows linearly with helix length, and a pitch dependent slope. The inset depicts the incoherent scattering from separate slabs of one turn in thickness.}
\label{fig7}
\end{figure}

If the polarizing power of a single turn of the helix is ${\bm P}_{unpol}=\text{Tr}[{\bm\sigma}\hat{\bf f}\hat{\bf f}^{\dagger}]/\text{Tr}[\hat{\bf f}\hat{\bf f}^{\dagger}]$ i.e. that given by Eq.\ref{polarization}, then the polarization through an additional turn of the helix (if the turns add up incoherently), can be computed from the density matrix $\rho^{(i+1)}=\hat{\bf f}(\mathds{1}+{\bm P}^{(i)}\cdot{\bm \sigma})\hat{\bf f}^{\dagger}/2$. Ignoring angular dependencies of incident electrons from one helix turn to the next, one obtains
\begin{widetext}
\begin{equation}
{\bm P}^{(i+1)}=\frac{1}{1+{\bm P}_{unpol}\cdot{\bm P}^{(i)}}\left({\bm P}_{unpol}+{\bm P}^{(i)}\frac{|h_0|^2-|{\bm h}|^2}{|h_0|^2+|{\bm h}|^2}+\frac{2 \Re\left[\left({\bm h}\cdot{\bm P}^{(i)}\right){\bm h}^{\dagger}\right]}{|h_0|^2+|{\bm h}|^2}-\frac{2 \Im\left[\left({\bm h}\times{\bm P}^{(i)}\right)h_0^{\dagger}\right]}{|h_0|^2+|{\bm h}|^2}\right ),
\end{equation}
\end{widetext}
where ${\bm P}^{(i)}$ is the polarization achieved after the $i$-th turn of the helix and ${\bm h}=(h_1,h_2,h_3)$(see Eq.\ref{finSigmas}).
Figure \ref{fig7} shows the polarization as a function of the number of helix turns for two different pitches smaller than the helix radius. We see a linear growth of the longitudinal polarization, proportional to the pitch size. The linear behavior, in spite of the complex relation between different components of the polarization, lends support for incoherent amplification. For larger pitch values, however, the relation between different components of the polarization can make the amplification non-monotonous, changing the polarization sign as the number of turns increase. The same mechanism can be envisaged for stacked lysine SAMs\cite{Naaman1999}. 

Summarizing, we have presented an elastic scattering scenario to describe spin-orbit interaction based polarization of electrons in chiral SAMs. The crux of our model is to consider multiple scattering off a helical model built from carbon atoms, disregarding bonding effects. The analytical approach, albeit consideration of only double scattering events, allows understanding the underpinnings of longitudinal polarization, the energy windows involved and the role of the chiral parameter. We have accounted for the strength of the SO
interaction from the microscopic  interactions by projecting the normalized plane waves onto the available atomic orbitals of the carbon scattering centers. This approach has no adjustable parameters and accounts directly for the SAM density.  Double and higher order scattering will be significant only within an energy window due to momentum transfer denominators that hinder the lateral scattering necessary to explore the target molecule. Finally,  the short coherence lengths for electrons at room temperature, lead to the scenario that the linear increase of the longitudinal polarization with chiral chain lengths\cite{NaamanDNA} is due to incoherent multiple forward scattering.

\section{Acknowledgments}
We acknowledge R. Naaman, Z. Helmud, D. Beratan, A. Nitzan and B. Berche for stimulating discussions, and EM thanks  IVIC for support under spintronics project 279. MR thanks the Chemistry Division of the NSF, for support under CHE-1058896.

\end{document}